\documentclass[
aps,%
12pt,%
final,%
notitlepage,%
oneside,%
onecolumn,%
nobibnotes,%
nofootinbib,% 
superscriptaddress,%
noshowpacs,%
centertags]%
{revtex4}

 \usepackage{multirow}

\usepackage{amsfonts,amsmath,amssymb,amsthm,nicefrac}

\usepackage[T1,T2A]{fontenc}
\usepackage[utf8]{inputenc}
\usepackage[russian,english]{babel}

\begin{document}
%\selectlanguage{english}

\title{Generalizing Stratonovich-Weyl axioms for  composite systems}% 

\author{\firstname{Arsen}~\surname{Khvedelidze}}
%  \\
%\email{akhved@jinr.ru}
\affiliation{A Razmadze Mathematical 
Institute, Tbilisi State University 
Tbilisi, Georgia}
\affiliation{Institute of Quantum Physics and Engineering Technologies,
Georgian Technical University 
Tbilisi, Georgia
}
\affiliation{Joint Institute for Nuclear Research, 
Dubna, Russia
}

%\date{\empty}

\begin{abstract}

The statistical model of quantum mechanics is based on  the  mapping   between operators on the Hilbert space and functions on the phase space. This map can be implemented by an  operator that satisfies  physically motivated Stratonovich–Weyl axioms. Arguments are given in favour of a certain extension of the axioms, provided that there is a priori  knowledge about the  composite nature  of the  quantum system. 
%The new postulate is formulated as requirement  that the partial tracing  of SW kernel of a whole system  with respect to one of its subsystems gives the SW kernel of the remaining  subsystem.
\end{abstract}

\maketitle

%\tableofcontents
%%%%%%%%%%%%%%%%%
\section{Introduction}
%%%%%%%%%%%%%%%%%%%%%%%%%%

Everyday experience tells us that a ``composite" is something admitting  partition into  an ``elementary''  which is  considered  to be primary and, therefore, not dividable any further. 
Moreover, it is assumed that, when  a ``composite'' object is divided into its  ``elementary'' parts,   all  properties of  the ``composite'' object  are determined by the characteristics of its constituents via certain rules of composition.
Several fundamental notions of classical physics originated from these intuitive assertions and were afterwards wrapped into rigorous mathematical concepts of measure space.
In particular, in classical statistical mechanics a vague idea of the  
``whole$\,\&$ part''  is embodied in the  measure-theoretic framework of the theory of probability where the ``composition'' rule  is encoded  in the \textit{principle of additivity of probability distributions},  one of  Kolmogorov's  axioms of the  probability space:\footnote{Under the probability space I assume the measure space $(\Omega, \mathcal{F}, \mathbb{P})$, such that the sample space (the space of all states of the system)  $\Omega$ is a
$2n\--$dimensional symplectic manifold, the space of events $\mathcal{F}\subset\Omega$  is represented by elements of $\sigma\--$algebra, and the probability measure 
$\mathbb{P}: \mathcal{F} \to [0,1]$ of event $A \in \mathcal{F}$ is given by the  Lebesgue integral, 
\(
\mathbb{P} (A)=\int_A\mu(dz)\rho(z),  
\)
with  the probability distribution function  $\rho(z) $.} 

\noindent{A.I. Non-negativity of the probability measure,  $\,\mathbb{P}(A)\geq 0; $}

\noindent{A.II. Finite  norm,  $\, \mathbb{P}(\Omega)=1; $}

\noindent{A.III. $\sigma\--$additivity for pairwise disjoint sets,
$\, \mathbb{P} (\cup_{i=1}^n\, A_i)= \sum_{i=1}^n\, P(A_i)\,. $
}

\noindent
The last axiom has a profound impact on the description of composite objects in classical theory.
Along with the assumption on the Boolean character of events, A.III.
results in an intuitively expected representation of the classical phase space of a composite system  in the form of the Cartesian product of the phase spaces of subsystems $A$ and $B$, 
\begin{equation}
\label{eq:Cartezian}
    \Omega_{A\times{B}}=\Omega_A \times \Omega_B\,.
\end{equation}
However,  even at the beginning of the quantum theory era, scientists realized that the aforementioned concepts would fail to describe a quantum world  (cf.\cite{BirkhoffNeumann1936,HealeyGomes2022,Seevinck2004}).  In particular, in the phase-space formulation of the quantum theory one cannot maintain  
the non-negativity axiom A.I  and  the composition rule  (\ref{eq:Cartezian})  as universal laws without encountering a contradiction
\cite{Wigner1932,Kirkwood1933,Dirac1945}. To avoid such a contradiction,  a  generalization of A.I-A.III was suggested through the replacement  of the $\sigma\-- $algebra by the non-Boolean lattice  which meant a fundamental transformation  of  conventional  Boolean logic into non-Boolean  quantum logic
(see e.g., \cite{BirkhoffNeumann1936,Elleman2018}).  As a result of  the change of the classical paradigm, the notion of probability distribution functions has been converted into the concept of quasiprobability distributions. 

In the present note, remaining in the framework of the phase-space  formulation
of quantum mechanics,  the analog of  (\ref{eq:Cartezian})  for  composite finite-dimensional quantum systems will be discussed. The  Stratonovich-Weyl (SW) axioms \cite{Stratonovich1957}  will be complemented  by  a new axiom of additivity for systems a priori known  to be compound ones.

%%%%%%%%%%%%%%%%%%%%%%%%%%%
\section{Composite quantum systems }

In quantum theory the analog of the Cartesian product (\ref{eq:Cartezian}) of the classical phase spaces of systems $A$  and $B$  is a  tensor product of the corresponding Hilbert spaces 
\(
{\mathcal H}_{AB} \subset  {\mathcal H}_{A}{\otimes }{\mathcal H}_{B}\,.
\)
The partial trace operation,
$
\mathrm{Tr}_{B}: \mathcal{H}_A\otimes \mathcal{H}_B \longmapsto \mathcal{H}_A\,, \
\mathrm{Tr}_{A}: \mathcal{H}_A\otimes \mathcal{H}_B \longmapsto \mathcal{H}_B\,
$,
allows us to extract information on each subsystem from  the density matrix $\varrho_{AB}$ of the whole system in the form  of the density matrices of subsystems  $\varrho_A$ and $\varrho_B:$ 
\begin{equation}
\label{eq:ptstate}
 \mathrm{Tr}_{B}: \quad \varrho{{}_{AB} }  \mapsto  \varrho_{A}:=\mathrm{Tr}_B \varrho{{}_{AB} }\,, 
 \qquad
  \mathrm{Tr}_{A}: \quad \varrho{{}_{AB} }  \mapsto  \varrho_{B}:=\mathrm{Tr}_A \varrho{{}_{AB} }
 \,.
\end{equation}
Aiming to construct the quasiprobability distributions of the subsystems for  a given composite system, we will propose the idea to use the partial trace operation (\ref{eq:ptstate}) not only for the states, but for their duals, the Stratonovich-Weyl kernels.

%%%%%%%%%%%%%%%%
\section{The Stratonovich-Weyl principles}
%%%%%%%%%%%%%

The Wigner quasiprobability distribution  $W_{\varrho}(\Omega_N)$ of an   
$N$\--dimensional  quantum system in a mixed state   is defined  by the pairing of density matrix $\varrho$ and   the Stratonovich-Weyl kernel $\Delta(\Omega_N)\,,$  which 
is defined on the  symplectic  space $\Omega_N:$ 
\begin{equation}
\label{eq:WF}
    W_{\varrho}(\Omega_N) = {\mbox{tr}(\varrho \, \Delta(\Omega_N))}\,. 
\end{equation}
The SW kernel $\Delta(\Omega_N)$ determines  a proper quasiprobability distribution provided the  following axioms  are satisfied 
(see e.g.,   
\cite{Stratonovich1957,AKh2021} and references therein):
\begin{enumerate}
\item[\bf 1.] {\bf Reconstruction} of state $\varrho$  by integrating the Wigner function over a phase space: 
\begin{equation}\label{eq:DMWigner}
\varrho =\int_{\Omega_N} \mathrm{d}\Omega_N\, \Delta(\Omega_N) W_\varrho(\Omega_N) \,;
\end{equation}	
\item[\bf 2.] {\bf Hermicity} of the SW kernel,  
\(\qquad
\Delta(\Omega_N)= \Delta(\Omega_N)^\dagger\,;
\)
\item[\bf 3.] {\bf Finite Norm} of a state given by the integral of the Wigner distribution: 
\begin{equation}
\mbox{tr}[ \varrho ]= \int_{\Omega_N} 
\mathrm{d}\Omega_N W_\varrho(\Omega_N)\,, 
\qquad
\int_{\Omega_N} \mathrm{d}\Omega_N\,\Delta(\Omega_N) = 1\,;
\end{equation}
\item[\bf 4.] {\bf Covariance}: The unitary transformation 
\(
\Delta^\prime(\Omega_N) =U(\alpha)^\dagger\Delta(\Omega_N)U(\alpha) 
\)
induces the symplectic change  
of coordinates $\boldsymbol{z}^\prime =
 T_{\alpha} \boldsymbol{z}\,, 
\, T_\alpha \in 
\mathrm{Sp}(d_N)\,, 
\boldsymbol{z}=\{z_1, z_2,\dots, z_{d_N}\} \in \Omega_N $.
\end{enumerate}
According to  \cite{AKh2021}, the axioms {\bf(1-4)} are  fulfilled  if
\begin{enumerate}
\item[(i)] The SW kernel $\Delta(\Omega_N)\,$ in (\ref{eq:WF})  is an element of the  dual space:
\begin{equation}
\label{eq:SWMaster}
\mathfrak{P}^\ast_N
=\{\, X \in M_N(\mathbb{C})\, |\, 
X=X^\dagger\,, {\mbox{tr}(\Delta(\Omega_N))=1}\,, \quad 
    {\mbox{tr}(\Delta(\Omega_N)^2)=N\,}
\}.
\end{equation}
The space of solutions to  (\ref{eq:SWMaster}), i.e., the moduli space,  is set by  an isotropy group, $\mathcal{P}_N := \mathfrak{P}^\ast_N/\mathrm{Iso_{{}_{SU(N)}}}(\Delta)$. For a regular $SU(N)\--$stratum  $\mathcal{P}_N$ is a polyhedron on $\mathbb{S}_{N-2}(1).$ 
\item[(ii)]The phase space $\Omega_N$ is
identified with the orbit 
$
\Omega_N = {SU(N)}/\mathrm{Iso_{{}_{SU(N)}}}(\Delta) \,. 
$
\end{enumerate}
Now, leaving {\bf(1-4)} in force, the {\bf 5-th} axiom on composite  quantum systems is proposed: 

{\bf 5.}  {\bf Composite systems axiom:}  \textit{Let   $\varrho_{AB}$ and  $\Delta(\Omega_N)$ be a density matrix  and a SW kernel  of composite system $\mathcal{H}_{A\times{B}}$, respectively, then
the Wigner functions of states $\varrho_{A}$ and $\varrho_B$
are constructed by pairing (\ref{eq:WF}) with partially reduced matrices 
$ \Delta_{A}:=\mathrm{Tr}_B\Delta \,\
 {and} \ \Delta_{B}:=\mathrm{Tr}_A\Delta
 $, i.e.,
$W_{\varrho_A}=\mathrm{Tr}\left(\varrho_{A}\Delta_{N_A}\right)\, and \ W_{\varrho_B}=\mathrm{Tr}\left(\varrho_{B}\Delta_{N_B}\right).
$
}

The extended system of axioms {\bf (1-5)} allows us to render  the following assertions:
\begin{itemize}
\item[(i).] The dual state space $\mathfrak{P}^\ast_{A\times{B}}$ of a binary composite system with $N_A\--$ and $N_B\--$ dimensional subsystems is $(N_A^2N_B^2-4)\--$dimensional  subspace of  $\mathfrak{P}^\ast_{N}$, defined as                                                        
\begin{equation}
\label{eq:SWMasterCom}
\mathfrak{P}^\ast_{A\times{B}}
=\{ X \in \mathfrak{P}^\ast _{N}\,,\,  N=N_AN_B\ | \ 
\mbox{Tr}_{{}_A}(\mbox{Tr}_{{}_{B}}
X)^2=N_{{}_A}\,, \quad  
\mbox{Tr}_{{}_B}(\mbox{Tr}_{{}_A}X)^2=N_{{}_B}
\}\,.
\end{equation}
 \item[(ii).] The phase space $\Omega_{A\times{B}}$ of a composite  system  is determined by the Local Unitary (LU)  group,
$\mathrm{LU}: = SU(N_A)\times SU(N_B) \subset{SU(N)}\,,$ corresponding 
to the  factorization  $\mathcal{H}_A\times \mathcal{H}_B$,  and by the isotropy group of the SW kernel:
$
\Omega_{A\times{B}} :={\mathrm{LU}}/{\mathrm{Iso_{{}_{LU}}}(\Delta)} \,;
$
\item[(iii).] The moduli space is the factor space
$
\mathcal{P}_{A\times{B}} :=
 {\mathfrak{P}^\ast_{A\times{B}}}/{\mathrm{LU}}\,.
 $
\end{itemize}
Below, the statements (i)\--(iii) are illustrated for a simplest binary system of a pair of qubits.

% % %
\section{Exemplifying a 2-qubit system}
%%%%%%%%%%%

 Let us consider the space of full rank $4\times4 $ density matrices $\varrho \in \mathfrak{P}_4$. Bearing in mind that the system is composed from 2-qubits, an adapted representation of a state is given  in terms of the   Bloch vectors of qubits  $\boldsymbol{\xi_A}$ 
 and 
$\boldsymbol{\xi_B}$  along with a  $3\times 3$ correlation matrix ${\mathcal{C}}:$
\begin{equation}
    \label{eq:2qDM}
 \varrho = \frac{1}{4}\,
\mathbb{I}_4 +\frac{\sqrt{6}}{4}\,
\left[
\boldsymbol{\xi_A}\cdot\boldsymbol{\sigma_A}
+
\boldsymbol{\xi_B}\cdot\boldsymbol{\sigma_B}
+ {\mathcal{C}}_{ij}\,
\sigma_i \otimes \sigma_j
\right]\,,   
\end{equation}
In (\ref{eq:2qDM}), $\boldsymbol{\sigma_A}=(\sigma_{10}, \sigma_{20}, \sigma_{30})$ and 
$\boldsymbol{\sigma_B}=(\sigma_{01}, \sigma_{02}, \sigma_{03})\,$
denote the elements of the Fano basis $\sigma_{\mu\nu}:=\sigma_\mu\otimes\sigma_\nu $ of the $\mathfrak{su}(4)$ algebra 
constructed out of  Pauli matrices $  \sigma_\mu=(\mathbb{I}_2, \sigma_1,\sigma_2, \sigma_3)\,.$ Similarly, vectors  $\boldsymbol{\eta_A}$ and $\boldsymbol{\eta_B}$ and a real  $3\times 3$ matrix ${\mathcal{E}}$ define the SW kernel $\Delta \in \mathfrak{P}_4$:
\begin{equation}
\label{eq:2qD}
\Delta= \frac{1}{4}\,
\mathbb{I}_4 +\frac{\sqrt{30}}{4}\,
\left[
\boldsymbol{\eta_A}\cdot\boldsymbol{\sigma_A}
+
\boldsymbol{\eta_B}\cdot\boldsymbol{\sigma_B}
+ {\mathcal{E}}_{ij}\,
\sigma_i \otimes \sigma_j
\right]\,.
\end{equation}
If a 4-level system is elementary, the master equations (\ref{eq:SWMaster}) impose the following condition:
\begin{equation}
\label{eq:ELC}
    \boldsymbol{\eta_A}^2+
\boldsymbol{\eta_B}^2+
\mathrm{tr}\left({\mathcal{E}}{\mathcal{E}}^T\right) = 1\,.
\end{equation}
However, if it is known that the system consists of two qubits,
according to (\ref{eq:SWMasterCom}), instead of (\ref{eq:ELC}) the individual norms of vectors $\boldsymbol{\eta_A}$, $\boldsymbol{\eta_B}$ and the matrix ${\mathcal{E}}$ are fixed:  
\begin{eqnarray}
\label{eq:LUconstraints}
\boldsymbol{\eta_A}^2=\frac{1}{10}\,,
\qquad
\boldsymbol{\eta_B}^2=\frac{1}{10}\,,
\qquad
\mathrm{tr}\left({\mathcal{E}}{\mathcal{E}}^T\right)=
\frac{4}{5}\,.
\end{eqnarray}
Therefore, the 2-qubit's dual space is $\mathfrak{P}^\ast_{2\times2}
\subset \mathfrak{P}^\ast_{4}$ such that $\dim(\mathfrak{P}^\ast_{2\times2})=15-3=12.$
The phase space $\Omega_4$, corresponding to an elementary 4-level system,  is one of the  $SU(4)\--$orbits \(\Omega_4= SU(4) / \mathrm{Iso_{{}_{SU(4)}}(\Delta)}\) on $ \mathfrak{P}_{4}$ .
When a 4-level system is considered to be a joint of  
2 qubits, then according to assertion (ii) the LU transformations and isotropy group $\mathrm{Iso}_{{}_{\rm LU}}(\Delta)$ of the SW kernel (\ref{eq:2qD}) define the phase space of 2-qubit $\Omega_{2\times 2}$, as well as its moduli space $\mathcal{P}_{2\times 2}$. In accordance with the 
2-qubit Hilbert space factorization, the LU transformations form the subgroup $ K=SU(2)\times SU(2) \subset SU(4)\,$ and the orbits of $K$ on $ \mathfrak{P}_{4}$ define $\Omega_{2\times2}\,.$
Hence, the issue  of describing the 2 qubit phase space is reduced to the  mathematical problem of classifying admissible types of  $K\--$orbits on $ \mathfrak{P}_{4}$.
Being restricted by the required volume of 
this note  I will illustrate the construction of a 2-qubit SW kernel  for only one class of  6-dimensional phase space
$\Omega_{2\times 2}$ and the moduli  
space 
$\mathcal{P}_{2\times 2}.$ 
In order to explicitly describe these spaces, it is convenient to decompose $g \in SU(4)$  into three   factors: 
\begin{equation}
\label{eq:SU4groupKAA'T}
g:= K\,\mathcal{A}\,T^3\,, \quad 
\mbox{with}\quad
\mathcal{A}=\exp{\mathfrak{a}}\exp{\mathfrak{a}^\prime}\,, \quad 
T^3 \--\mbox{maximal~torus}\,. 
\end{equation}
In (\ref{eq:SU4groupKAA'T})
$\mathfrak{a}$ and $\mathfrak{a}^\prime$ are the Abelian subalgebras 
in the direct sum decomposition of the algebra:
\begin{equation}
\label{eq:NewDecomp}
  \mathfrak{su}(4) = \mathfrak{k}
  \oplus  \mathfrak{a}\oplus
\mathfrak{a}^\prime\oplus \mathfrak{k}^\prime\,,
\qquad 
\mathfrak{k}:=\mathfrak{su}(2)\oplus \mathfrak{su}(2)\,, \quad 
  \mathfrak{k}^\prime:=
  \mathfrak{u}(1)\oplus \mathfrak{u}(1)
   \oplus{u}(1)\,,
\end{equation}
such that  the 
following commutator relations  hold:
\footnote{
Note that if $
\{\lambda_1,\dots,\lambda_{15}\}
=\frac{i}{{2}}\{\sigma_{10},
\sigma_{20}, \sigma_{30}, \sigma_{01}, \sigma_{02}, \sigma_{03}, \sigma_{11},
\sigma_{12}, \sigma_{13}, \sigma_{21}, \sigma_{22}, \sigma_{23},
\sigma_{31}, \sigma_{32}, \sigma_{33}\},
$ then 
$
\mathfrak{a}=\mbox{span}\{\lambda_{11}\,,
\lambda_9\, \lambda_{13}\}, \,
\mathfrak{a}^\prime=\mbox{span}\{\lambda_{4}\,,
\lambda_1\, \lambda_{7}\}\,, $  while the  torus 
algebra is
$\mathfrak{k}^\prime:=
\mbox{span}\{\lambda_{3}\,,
\lambda_6\, \lambda_{15}\}\,, $ and the algebra of subgroup $K$  reads
$\mathfrak{k}
:=
\mbox{span}\{-\lambda_{14}\,,
\lambda_2\,, -\lambda_{8}\,; -\lambda_{5}\,,
\lambda_{12}\,, -\lambda_{10}\}\,.
$} 
\begin{eqnarray}
&&[\mathfrak{a}^\prime\,, \mathfrak{a}] \subset \mathfrak{l}\,,\qquad
[\mathfrak{k}, \mathfrak{k}] \subset \mathfrak{k}\,, \qquad 
[\mathfrak{k}^\prime, \mathfrak{k}^\prime ] \subset \mathfrak{k}^\prime\,,
\qquad [\mathfrak{k}, \mathfrak{k}^\prime] \subset \mathfrak{a}\oplus
\mathfrak{a}^\prime\,.
\end{eqnarray}
Using  (\ref{eq:SU4groupKAA'T})
for unitary factor $U$ in SVD of
$4\Delta_4=U\left(\mathbb{I}_4+ \sqrt{30}\,\sum_{\lambda_\alpha\in\mathfrak{k}^\prime}\,\lambda_\alpha \mu_\alpha \right)U^\dagger\,, 
$ one can be convinced that the master equations (\ref{eq:LUconstraints}) describe a bundle  of a unit 
2-sphere and two ellipsoids $\mathrm{E_A}$ and $\mathrm{E_B}$ in the space $\mathcal{P}_4 $ with the Cartesian  coordinates  $\boldsymbol{\mu}=\{\mu_3, \mu_6, \mu_{15}\}$: 
\begin{equation}
\label{eq:LUconstraintsMu}
 \boldsymbol{\mu}\boldsymbol{\mu}^T=1\,, \qquad 
\mathrm{E_A}:\quad
\boldsymbol{\mu}\mathbb{A}\boldsymbol{\mu}^T= 1\,, \qquad
\mathrm{E_B}:\quad 
\boldsymbol{\mu}\mathbb{B}\boldsymbol{\mu}^T = 1\,,
\end{equation} 
The $3\times 3$ matrices $\mathbb{A}$ and $\mathbb{B}$ in (\ref{eq:LUconstraintsMu})
are  constructed out of the adjoint matrix $\mathcal{A}\lambda_\nu \mathcal{A}^\dagger = O_{\nu \mu}\, \lambda_\mu
$:
\begin{equation}
  \mathbb{A}_{\alpha\beta}: = \frac{4}{3} \sum_{i=1,2,3}O_{\alpha i}O^T_{i\beta}
 \,, \qquad  
 \mathbb{B}_{\alpha\beta}: = \frac{4}{3} \sum_{i=4,5,6}O_{\alpha i}O^T_{i\beta}
 \,.  
\end{equation}
The moduli space $\mathcal{P}_{2\times2}$ of a 2-qubit system  is  determined
from analyzing the following  pairwise characteristic polynomials of ellipsoids $\mathrm{E_A}\,, \mathrm{E_B}$ and a unit 2-sphere: 
\begin{equation}
\label{eq:cheq2qubit}
f_{{}_{\mathrm{E_A}\cap\, \mathbb{S}_2}}(t) =\det(t\mathbb{I}_3 + \mathbb{A}),\quad
f_{{}_{\mathrm{E_B}\cap\, \mathbb{S}_2}}(t)=\det(t\mathbb{I}_3 + \mathbb{B}), \quad 
f_{{}_{\mathrm{E_A}\cap\,
\mathrm{E_B}}}(t) =\det(t\mathbb{A} + \mathbb{B})
\,.
\end{equation}
According  to  \cite{Wang2001}, each 
characteristic polynomial in (\ref{eq:cheq2qubit}) always has at least one negative root. Moreover, the ellipsoids and the 2-sphere overlap iff  characteristic  polynomials 
have no positive roots. More on the geometric properties of 
$\mathcal{P}_{2\times2}$ will be given elsewhere.

%%%%%%%%%%%%%%%%%%%% SUMMARY
\section{Summary}
 %%%%%%%%%%%%%%%%%%%%%%%%%%%%%   

Discussing the phase\--space approach to elementary vs composite quantum systems, it is worth emphasizing the common principle used: the underlying symmetry of a system  dictates the construction of the basic quantities in both cases. In the first case, the  global $SU(N)$ symmetry sets the  phase space 
$\Omega_N:=SU(N)/\mathrm{Iso{{}_{SU(N)}}}(\Delta),$ as well as the moduli space  $\mathcal{P}_N,$ via the master equations (\ref{eq:SWMaster})  on the  orbit space $\mathfrak{P}_N^\ast/SU(N).$ In the second case,  dealing with another pair of spaces $(\mathfrak{P}_{A\times B}, \mathfrak{P}^\ast_{A\times{B}}),$  the local sub-symmetry, 
$LU:=SU(N_A)\times SU(N_B) \subset SU(N),$ comes into play once more. The latter defines  the phase space $\Omega_{N_A\times N_B}$   
 and the moduli space 
$\mathcal{P}_{N_A\times N_B}$ via  
the extended master equations (\ref{eq:SWMasterCom}) on  the orbit  space $\mathfrak{P}_{A\times{B}}^\ast/LU $.

%%%%%%%%%%%%%%%%
\begin{acknowledgments}
The work is supported in part by  the Shota Rustaveli National Science Foundation of Georgia, Grant FR-19-034. 
\end{acknowledgments}
%%%%%%%%%%%%%%%%

%%%%%%%%%%%%%%%%%%%%%% Bibliography 


\begin{thebibliography}{99}
%1 
\bibitem{BirkhoffNeumann1936}
G.~Birkhoff and J.von~Neumann,
%The logic of quantum mechanics,
Annals Math. {\bf 37}, 823,
%-843, 
(1936).
%2
\bibitem{HealeyGomes2022}
R.~Healey and H.~Gomes, "Holism and Nonseparability in Physics", The Stanford Encyclopedia of Philosophy (Spring 2022 Edition), Edward N. Zalta (ed.). 
%URL = https://plato.stanford.edu/archives/spr2022/entries/physics-holism.
%
%\bibitem{Maudlin1998}T.~Maudlin, Part and whole in quantum mechanics, in  Interpreting Bodies, Elena Castellani (ed.), Princeton University Press.  46--60 (1998)
%
% 3
\bibitem{Seevinck2004}M.P.~Seevinck, 
%Holism, physical theories and quantum mechanics, 
Studies in History and Philosophy of Modern Physics {\bf 35},
693,
%–712
(2004).
% 4
\bibitem{Wigner1932}
E. Wigner, 
%On the quantum correction for thermodynamic equilibrium, 
Phys. Rev. \textbf{40}, 749 (1932).
% 5
\bibitem{Kirkwood1933}
J. G. Kirkwood, 
%Quantum statistics of almost classical assemblies, 
Phys. Rev. \textbf{44}, 31 (1933).
% 6
\bibitem{Dirac1945}
P.A.M.~Dirac, 
%On the analogy between classical and quantum mechanics, 
Rev. Mod. Phys. \textbf{17}1, 195, (1945).
% 7
\bibitem{Elleman2018}
D.~Ellerman,
%Logical Entropy: Introduction to Classical and Quantum Logical Information Theory. 
Entropy  2018 Sep 6; \textbf{20}, 679. (2018).
% 8
\bibitem{Stratonovich1957}
R.L.~Stratonovich,
%On distributions in representation space,
Sov.Phys.JETP \textbf{4},  891
%-898 
(1957).
% 9
\bibitem{AKh2021}V.~Abgaryan  and A.~Khvedelidze, 
%{On families of Wigner functions for $N$-level quantum systems}, 
Symmetry {\bf 13}, 1013, (2021).
% 10
\bibitem{Wang2001}W.~Wang, \emph{et al.},
%An algebraic condition for the separation of two ellipsoids,
Computer Aided Geometric Design,
\textbf{18}, 531,
%-539
(2001).
\end{thebibliography}
\end{document}